\documentclass[aps,preprint]{revtex4}
\usepackage{amssymb}
\usepackage{epsfig}
\usepackage{color}
\usepackage{amsmath, amsthm}
\usepackage{amsmath, amsfonts}
\usepackage{amsmath}
\usepackage[all]{xy}

\makeatletter

\newcommand{\Rmnum}[1]{\expandafter\@slowromancap\romannumeral #1@}
\makeatother

\newcommand{\pa}{\partial}

\newcommand{\td}{\tilde}

\begin{document}
%\begin{CJK*}{GBK}{song}
%\draft
\preprint{USTC-ICTS-09-13}
\title{
Hydrogen Atom  and Time Variation of Fine-Structure Constant}
\author{Mu-Lin Yan}
 \email{mlyan@ustc.edu.cn}

\affiliation{Interdisciplinary Center for Theoretical Study,
Department of Modern Physics, University of Science and Technology
of China, Hefei, Anhui 230026, China}

\date{}
%\maketitle
\begin{abstract}
\noindent In this paper, we have solved the de Sitter special
relativistic ($\mathcal{SR}_{cR}$-) Dirac equation of hydrogen in
the earth-QSO(quasar) framework reference by means of the adiabatic
approach. The aspects of geometry effects of de Sitter space-time
described by Beltrami metric are explored and taken into account. It
is found that the $\mathcal{SR}_{cR}$-Dirac equation of hydrogen is
a time dependent quantum Hamiltonian system. We provide an explicit
calculation to justify the adiabatic approach in dealing with this
time-dependent system. Since the radius of de Sitter sphere $R$ is
cosmologically large,  the evolution of the system is very slow so
that the adiabatic approximation legitimately works with high
accuracy. We conclude  that   the electromagnetic fine-structure
constant, the  electron mass and the Planck constant are time
variations. This prediction of fine-structure constant is consistent
with the presently available observation data. For confirming it
further, experiments/observations are required.

\vskip0.2in

\noindent PACS numbers: 03.30.+p; 03.65.Ge; 04.62.+v; 95.85.Sz.\\
Key words: Hydrogen atom, Time variation of fine structure constant,
de-Sitter invariance, Special Relativity.

\end{abstract}

\maketitle

\begin{center}
Contents
\end{center}
\begin{enumerate}
\item Introduction
\item Review of the classical mechanics for free particle in de Sitter special
relativity $\mathcal{SR}_{cR}$

\begin{enumerate}
\item The Lagrangian-Hamiltonian formalism
\item Space-time symmetry of de Sitter special relativity and
the Neother charges
\end{enumerate}

\item Quantum mechanics  in de Sitter special relativity
\item Hydrogen atom in earth-QSO reference frame and variation of
electromagnetic fine-structure constant

\begin{enumerate}
\item $\mathcal{SR}_{cR}$-Dirac equation for hydrogen atom on QSO
\item Solution of usual $\mathcal{SR}_c$-Dirac
equation for hydrogen atom at QSO
\item Beltrami-geometry effects  in  $\mathcal{SR}_{cR}$-Dirac
equation
\item $\mathcal{SR}_{cR}$-Dirac
equation for  spectra of hydrogen
\item Adiabatic approximation solution to $\mathcal{SR}_{cR}$-Dirac
spectra equation
\item Comparing theory predictions to observations
\end{enumerate}
\item Summery and discussions
\end{enumerate}

 Appendix A: Electric Coulomb Law in QSO-Light-Cone Space

 Appendix B: Adiabatic approximative wave functions in
$\mathcal{SR}_{cR}$-Dirac equation of

hydrogen

References

\newpage

\section{Introduction}
\noindent The life time of a stable atom, e.g., the hydrogen atom,
is almost infinitely long. We can practically compare the spectra of
atoms at nowadays laboratories  to ones emitted from the atoms  of a
distant galaxy. The time interval could be on the cosmic scales.
Such observation of spectra of distant astrophysical objects may
encode some cosmologic information in the atomic energy levels at
the position and time of emission. During last decade, several
interesting experiments based on this idea in principle were
reported in literature, and the fine-structure constant
$\alpha$-variation in the absorption spectra of quasi-stellar
objects (QSOs) were observed (see incomplete list of reference
\cite{webb06,Murphy03b,Murphy03a,murphy01b,murphy01a,webb01,webb99}
, and also the review articles \cite{uzan}\cite{rev2}\cite{Fritzsch}
and the references within). In order to reveal the probable physics
behind these experimental discoveries, we strongly suggest to
reexamine the dynamic theories of atoms, typically of the hydrogen,
with both fine-structure effects and cosmological effects taken into
account. As is well known, in the ordinary relativistic quantum
mechanics based on Einstein¡¯s Special Relativity (denoted  as
$\mathcal{SR}_c$ hereafter), the hydrogen's fine-structure spectra
are independent of any cosmologic effects. For instance, the atomic
spectra in this theory do not change in time due to the fact that
 the Harmiltonian $H_c$ is time independent. Therefore, investigations
 of quantum theory of atoms  at cosmic
scale should be based on some extension of Einstein's special
relativity $\mathcal{SR}_c$ in which the flat Minkowski spacetime is
replaced by de Sitter spacetime, and of course it should be a
challenge.

A natural extension of  $\mathcal{SR}_c$  is the de Sitter invariant
Special Relativity (or the Special Relativity in space-time of a
positive constant curvature $1/R$). By requiring the law of inertia
for free particles to be true in the de Sitter special relativity,
authors of \cite{look}\cite{Lu74} found out that the space-time
geometry is described by  Beltrami metric (instead of usual Lorentz
metric), and the space-time coordinate transformations to preserve
Beltrami metric exists (see \cite{Yan1} for an English version).
Thus the de Sitter special relativity was formulated in
\cite{look}\cite{Lu74}. In the recent years, there have been further
studies on this theory in \cite{Guo1}\cite{Yan1}\cite{Yan2}. There
is one universal parameter $c$ (speed of light) in the Einstein's
Special Relativity $\mathcal{SR}_{c}$. By contrast, there are two
universal parameters in the de Sitter Special Relativity:
 $c$ and $R$ (the radius of de Sitter
sphere and to character the cosmic radius). So, we will  denote
latter shortly as
  $\mathcal{SR}_{cR}$  hereafter.
In \cite{Yan1}, the Hamiltonian formalism of de Sitter special
relativity was developed. In \cite{Yan2}, by requiring that the
results of time-variation of fine structure constant  in the
absorption spectra of QSOs in
\cite{webb06,Murphy03b,Murphy03a,murphy01b,murphy01a,webb01,webb99}
are consistent with ones measured in Oklo nature fission reactor
\cite{Oklo}, the $R$ was estimated to be $10^{11}ly$ to $ 10^{12}ly$
approximately.

Similar to classical $\mathcal{SR}_{c}$ mechanics, the
Lagrangian-Hamiltonian formulation  of $\mathcal{SR}_{cR}$ mechanics
are self-consistent, and have been appropriately established in
\cite{Yan1}. The most significant difference between these two
theories are that the free particle's Hamiltonian for
$\mathcal{SR}_{c}$ is spacetime independent $H_c=H_c(p_i)$, while
for $\mathcal{SR}_{cR}$ the Hamiltonian is
\begin{equation}\label{1}
H_{cR}=H_{cR}({c^2t^2\over R^2},\; {(x^i)^2\over R^2},\; \pi_i)
\end{equation}
which depends on time explicitly. Of course, when $R\rightarrow
\infty$, $ H_{cR}({c^2t^2\over R^2},\; {(x^i)^2\over R^2},\;
\pi_i)\rightarrow H_c(p_i)$. In $\mathcal{SR}_{cR}$ the particle's
conserved energy $E$ and momenta $p^i$ are different from its
canonical energy (or Hamiltonian $H_{cR}$) and canonical momenta
$\pi_i$. $E$ and $p^i$ appear as the Noether charges of the de
Sitter symmetry for space-time of $\mathcal{SR}_{cR}$ mechanics
\cite{Yan1}. The quantization of such a system is obviously
nontrivial. Different from the quantization of both  Newtonian
mechanics and of  $\mathcal{SR}_c$ mechanics, the operator ordering
 of $``x"$ and $``\rm{momentum}"$ must be taken into
account for free particle motions in $\mathcal{SR}_{cR}$-quantum
mechanics \cite{Yan1}. It has been shown that the Weyl ordering is
necessary for protecting the isometry symmetries $SO(1, 4)$ of de
Sitter spacetime, and the wave-equation of spinless particle was
shown to be the Klein-Gordon (KG) equation in de Sitter space-time
with Beltrami metric \cite{Yan1}. In the present paper, we will base
on such  KG equation to construct the $\mathcal{SR}_{cR}$-Dirac
equation for spin 1/2 particles. Namely, the tetrad and the spin
connection corresponding to the Beltrami metric will be derived.
Nextly, by treating the Coulomb electric interaction between nucleus
and electron as $U(1)$-EM gauge potential and basing on gauge
covariant principle, we finally obtain the $\mathcal{SR}_{cR}$-Dirac
equation for electron in hydrogen atom.

The main purpose of this paper is to study the spectra of hydrogen
atom located on distant astrophysical objects, e.g., QSO. Because
people gets astrophysics information by observation, all observable
optic objects must be on the light cone of the earth. Thus, we
should solve the $\mathcal{SR}_{cR}$-Dirac equation for electron in
hydrogen atom in the earth-QSO reference frame, whose origin is at
earth and QSO locates on the light cone. The geometry is determined
in Beltrami metric of de Sitter space-time. It is expected that the
solutions will show both effects of fine-structure and effects of
cosmology in the spectra of such hydrogen atom. To do so, we have to
solve time-dependent Harmiltonian problem due to (\ref{1}) in
quantum mechanics. Our explicit calculations show that since $R$ is
cosmologically large and $R>>ct$, factor $ (c^2t^2/ R^2)$ makes the
time-evolution of the system be so slow that the adiabatic
approximation \cite{Born} will  legitimately
 works.

Generally, to a $H(x,t)$, we may express it as
$H(x,t)=H_0(x)+H'(x,t)$. Suppose two eigenstates $|s\rangle$ and $|m
\rangle$ of $H_0(x)$ do not generate, i.e., $\Delta E\equiv \hbar (
\omega_{m}-\omega_s)=\hbar \omega_{ms}\neq 0$.
 The validness of for  adiabatic
approximation relies on the fact that the variation of the potential
$H'(x,t)$ in the the Bohr time-period $(\Delta
T_{ms}^{(Bohr)})\dot{H}'(x,t)_{ms}=(2\pi/\omega_{ms})\dot{H}'(x,t)_{ms}$
is much less than $\hbar \omega_{ms}$. That  makes the quantum
transition from  state $|s\rangle$ to  state $|m \rangle$ almost
impossible. Thus,  the non-adiabatic effect corrections are small
enough (or tiny) , and  the adiabatic approximations are proper
\cite{Messian}.

For adiabatic quantum system, the states are quasi-stationary in all
instants, and hence the time variable becomes a parameter in
Hamiltonian. In this approximation approach, the time-dependent
Hamiltonian system was reduced to a system with time-parameter
(rather than a time-dynamic variable), and then the problem becomes
handleable and solvable approximatively.

By means of adiabatic approximation approach, we solve the
stationary $\mathcal{SR}_{cR}$-Dirac equation for hydrogen atom, and
the spectra of the corresponding Hamiltonian with time-parameter are
obtained. As a result, we find out that the fine structure constant
and the mass of electron vary  as cosmic time going by. This is a
interesting  consequence of the theory. We will compare the
prediction of our theory with the observation data of
\cite{webb06,Murphy03b,Murphy03a,murphy01b,murphy01a,webb01,webb99}
in the end of the paper. It will be pointed out that the prediction
is in agreement with the observation.

$\mathcal{SR}_{cR}$-quantum mechanics for atom could be thought of
as a cosmological atom physics theory. Since the works in this field
would be helpful to reveal information about atomic energy levels of
emission from cosmological distant object, the results and
predictions could be interesting. In addition, the studies on
$\mathcal{SR}_{cR}$ belong to examining the base of the relativity
theory from its beginning. It would be also meaningful to search
what physics effects could distinguish the predictions of
$\mathcal{SR}_{cR}$ from ones of $\mathcal{SR}_{c}$. The results of
this paper may indicate that the special relativity for cosmologic
large space-time scale may be beyond the Einstein's special
relativity $\mathcal{SR}_c$.

The contents of the paper are organized as follows: In section II,
we recall the classical mechanics of de Sitter special relativity;
Section III is devoted to discuss the quantum mechanics of de Sitter
special relativistic. The quantum wave equations  both of sipinless
particle and of spin-1/2 one are constructed; Section IV is the
major part of the paper. In this section, we solve the
$\mathcal{SR}_{cR}$-Dirac equation for hydrogen atom step by step:
to derive $\mathcal{SR}_{cR}$-Dirac equation for hydrogen atom on
QSO; to discuss Solution of usual $\mathcal{SR}_c$-Dirac equation
for hydrogen atom at QSO, and the Beltrami-geometry effects  in
$\mathcal{SR}_{cR}$-Dirac equation; and then
$\mathcal{SR}_{cR}$-Dirac equation for  spectra of hydrogen is
obtained and further solved by means of adiabatic approximation. In
the end of this section we exhibit  that the fine structure constant
is time-variation and compare it with the observations; Finally, we
briefly summarize and discuss the results of the paper. In Appendix
A, we derive the electric Coulomb Law in QSO-Light-Cone Space; In
Appendix B, we show the calculations of adiabatic approximative wave
functions in $\mathcal{SR}_{cR}$-Dirac equation of hydrogen in
detail.

\section{REVIEW OF THE classical mechanics for free particle in de Sitter special
relativity \cite{Yan1}}

\subsection{The Lagrangian-Hamiltonian formalism }
\noindent We begin with a brief review of the classical mechanics
for a free particle in de Sitter special relativity. The Lagrangian
 is
\begin{equation}\label{27}
 L_{{cR}}=-m_0c{ds\over
 dt}=-m_0c{\sqrt{B_{\mu\nu}(x)dx^\mu dx^\nu}\over dt}=-m_0c{\sqrt{B_{\mu\nu}(x)\dot{x}^\mu \dot{x}^\nu}},
 \end{equation}
where $\dot{x}^\mu=\frac{d}{dt}x^\mu$, $B_{\mu\nu}(x)$ is Beltrami
metric\cite{look,Lu74,Guo1,Yan1,Yan2}:
\begin{eqnarray}\label{star28}
B_{\mu\nu}(x)={\eta_{\mu\nu} \over \sigma (x)}+{1\over R^2
\sigma(x)^2} \eta_{\mu\lambda}\eta_{\nu\rho} x^\lambda
x^\rho,~~~{\rm{with}}~~~\sigma(x)\equiv 1-{1\over R^2}
\eta_{\mu\nu}x^\mu x^\nu,
\end{eqnarray}
and $R$ which is assumed to be a fundamental constant in
$\mathcal{SR}_{cR}$ stands for  the radius of the pseudo-sphere in
{\it dS}-space. Setting up the time $t=x^0/c$,   $B_{\mu\nu}(x)$
becomes
\begin{eqnarray}\label{star29}
ds^2&=&B_{\mu\nu}(x) dx^\mu dx^\nu
=\widetilde{g}_{00}d(ct)^2+\widetilde{g}_{ij}\left[(dx^i+N^id(ct))
(dx^j+N^jd(ct))\right]\\ \nonumber &=& c^2 (dt)^2
\left[\widetilde{g}_{00} +\widetilde{g}_{ij}({1\over
c}\dot{x}^i+N^i) ({1\over c}\dot{x}^j+  N^j)\right],
\end{eqnarray}
where
\begin{eqnarray}\label{star30}
\widetilde{g}_{00}&=&{R^2\over \sigma(x) (R^2-c^2t^2)},\\
\label{star31} \widetilde{g}_{ij}&=&{\eta_{ij}\over \sigma (x)}+
{1\over
R^2\sigma(x)^2}\eta_{il}\eta_{jm}x^lx^m,\\
\label{star32} N^i&=&{ctx^i \over R^2-c^2t^2}.
\end{eqnarray}
By substituting eqs.(\ref{star28})--(\ref{star32}) into (\ref{27}),
we recast the Lagrangian as
\begin{equation}\label{33}
 L_{{cR}}=-m_0c^2 \sqrt{\widetilde{g}_{00} +\widetilde{g}_{ij}({1\over
c}\dot{x}^i+N^i) ({1\over c}\dot{x}^j+  N^j)}.
 \end{equation}
from which the following identity results in,
\begin{equation}\label{initial motion condition}
 {\pa L_{{cR}}\over \pa x^i}= {\pa^2 L_{{cR}}\over \pa t \pa \dot{x}^i}
+ {\pa^2 L_{{cR}}\over \pa x^j \pa \dot{x}^i}\dot{x}^j.
 \end{equation}
Considering the Euler-Lagrangian equation
\begin{equation}\label{15}
{\pa L_{cR}\over \pa x^i}={d\over dt}{\pa L_{cR}\over \pa
\dot{x}^i},
\end{equation}
we obtain the solution of equation of motion for free particle :
\begin{equation}\label{38}
\ddot{x}^j=0, ~~ \dot{x}^j={\rm constant}.
\end{equation}
Next step is to derive the canonic momenta and the canonic energy
(i.e., Harmiltonian). By the eq.(\ref{33}), they reads
\begin{eqnarray}\label{40}
   \pi_{i} &=&\frac{\pa L_{cR}}{\pa
\dot{x}^i} = -m_0 \sigma(x) \Gamma B_{i \mu}\dot{x}^{\mu} \\
 \label{40a}    H_{cR}
&=&\sum_{i=1}^3 \frac{\pa L_{cR}}{\pa \dot{x}^i} \dot{x}^i
-L_{cR}=m_0 c \sigma(x) \Gamma B_{0 \mu}\dot{x}^{\mu}.
  \end{eqnarray}
where {\begin{eqnarray} \label{new parameter}
 \Gamma^{-1}\hskip-0.1in =\sigma(x) \frac{ds}{c dt}={1\over R} \sqrt{(R^2-\eta_{ij}x^i
x^j)(1+\frac{\eta_{ij}\dot{x}^i \dot{x}^j}{c^2})+2t \eta_{ij}x^i
\dot{x}^j -\eta_{ij}\dot{x}^i \dot{x}^j t^2+\frac{(\eta_{ij}
x^i\dot{x}^j)^2}{c^2}}.
\end{eqnarray}}
Under the equation of motion Eq.(\ref{38}), we have the following
relation
\begin{equation}\label{38a} \dot{\Gamma}|_{\ddot{x}^i=0}=0,
\end{equation}
whose corresponding one in $\mathcal{SR}_c$ is
\begin{equation}\label{38b}
\dot{\gamma}|_{\ddot{x}^i=0}\equiv{d\over dt}\left. \left({1\over
\sqrt{1-v^2/c^2}}\right)\right|_{v={\rm constant}}=0.
\end{equation}
It is easy  to check that
\begin{equation}\label{38c}
\lim_{R\rightarrow \infty} \Gamma= \lim_{x^i\rightarrow 0} \Gamma
=\gamma\equiv \frac{1}{\sqrt{1-\frac{v^2}{c^2}}}.
\end{equation}
And, in the $R \to \infty$ limit, $\pi_i$ and $H_{cR}$ go back to
the standard Einstein Special Relativity's expressions:
\begin{equation} \label{44}
 \pi_{i}|_{R\rightarrow \infty}= \frac{m_0
 v_i}{\sqrt{1-\frac{v^2}{c^2}}},~~~~
    H_{cR}|_{R\rightarrow \infty} =\frac{m_0 c^2}
{\sqrt{1-\frac{v^2}{c^2}}}.
 \end{equation}
where $v_i=-\eta_{ij}\dot{x}^j$. In the Table I, we listed some
results of Lagrange formulism  both in the ordinary special
relativity $\mathcal{SR}_c$ and in the de Sitter invariant special
relativity $\mathcal{SR}_{cR}$. Comparing the results in
$\mathcal{SR}_{cR}$ with ones in well known $\mathcal{SR}_{c}$, we
learned that as an extending theory of $\mathcal{SR}_{c}$,
$\mathcal{SR}_{cR}$ can simply be formulated by a variable
alternating in $\mathcal{SR}_{c}$: 1) $\eta_{\mu\nu}\Rightarrow
B_{\mu\nu}$; 2) $\gamma \Rightarrow \sigma \Gamma$. This is a
natural and nice feature for the Lagrangian formulism of
$\mathcal{SR}_{cR}$.

\begin{table}[hptb]
\begin{center}
\caption{Metric, Lagrangian, equation of motions, canonic momenta,
and Hamiltonian in the special relativity, $\mathcal{SR}_c$, and in
the de Sitter special relativity, $\mathcal{SR}_{cR}$. The
quantities $\gamma^{-1}=\sqrt{1+{\eta_{ij}\dot{x}^i \dot{x}^j \over
c^2}} $ and $\Gamma^{-1}\hskip-0.1in ={1\over R}
\sqrt{(R^2-\eta_{ij}x^i x^j)(1+\frac{\eta_{ij}\dot{x}^i
\dot{x}^j}{c^2})+2t \eta_{ij}x^i \dot{x}^j -\eta_{ij}\dot{x}^i
\dot{x}^j t^2+\frac{(\eta_{ij} x^i\dot{x}^j)^2}{c^2}} $ (see
eq.(\ref{new parameter})).}
\begin{tabular}{|c|c|c|}\hline\hline
    & $\mathcal{SR}_c$ & $\mathcal{SR}_{cR}$ \\ \hline
space-time metric & $\eta_{\mu\nu}$ & $B_{\mu\nu}(x),$
(Eq.(\ref{star28})) \\ \hline Lagrangian & $L_c=-m_0c^2\gamma^{-1}$
& $L_{cR}=-m_0c^2\sigma^{-1}\Gamma^{-1}$ \\ \hline equation of
motion & $v^i=\dot{x}^i=$constant, ( or $\dot{\gamma}=0$)
& $v^i=\dot{x}^i=$constant, ( or $\dot{\Gamma}=0$) \\
\hline canonic momenta & $\pi_i=-m_0\gamma \eta_{i\mu}\dot{x}^\mu $
& $\pi_i=-m_0\sigma \Gamma B_{i\mu}\dot{x}^\mu $ \\ \hline
Hamiltonian & $H_c=m_0c\gamma \eta_{0\mu} \dot{x}^\mu$ &
$H_{cR}=m_0c\sigma \Gamma B_{0\mu} \dot{x}^\mu$ \\ \hline \hline
\end{tabular}
\end{center}
\end{table}
Combining  Eq.(\ref{40}) with Eq.(\ref{40a}), the covariant
4-momentum in $\cal{B}$ is:
\begin{equation}\label{46-1}
\pi_{\mu}\equiv (\pi_0,\;\pi_i)=(-\frac{H_{cR}}{c},\pi_i)=-m_0
\sigma \Gamma B_{\mu \nu} \dot{x}^{\nu}= -m_0 cB_{\mu
\nu}{d{x}^{\nu}\over ds},
\end{equation}
and
\begin{eqnarray}\label{dispersion}
B^{\mu\nu}\pi_{\mu}\pi_{\nu}&=&m_0^2 c^2.
\end{eqnarray}
From eqs.(\ref{33}) (\ref{40}) (\ref{40a}) (\ref{dispersion}), we
have the standard form of $H_{cR}(t,x^i,\pi_i)$ as follows
\begin{eqnarray}\label{hamilton}
H_{cR}=\sqrt{\widetilde{g}_{00}}\sqrt{m_0^2c^4
-c^2\widetilde{g}^{ij}\pi_i\pi_j}-c\pi_iN^i,
\end{eqnarray}
where $\widetilde{g}_{00},\;N^i$ have been shown in
eqs.(\ref{star30}) (\ref{star32}), and
$\widetilde{g}^{ij}={\sigma(x)(\eta^{ij}-\frac{x^i x^j}{R^2-c^2
t^2})}$ from eq.(\ref{star31}). It is straightforward to to get the
following canonical equations
\begin{eqnarray}\label{canonical equation}
\begin{array}{c}
     \dot{x}^i = \frac{\partial H_{cR}}{\partial \pi_i}=\{H_{cR}, x^i\}_{PB}  \\
     \dot{\pi}_i =-\frac{\partial H_{cR}}{\partial
x^i}=\{H_{cR}, \pi_i\}_{PB}.
  \end{array}
\end{eqnarray}
where the Poisson bracket
\begin{eqnarray}\label{Poisson}
\{x^i, \pi_j\}_{PB}=\delta^i_j, \;\;\{x^i, x^j\}_{PB}=0,\;\;\{\pi_i,
\pi_j\}_{PB}=0
\end{eqnarray}
 are as usual. It is also straightforward to check $\dot{x}^i=constant$ by
eq.(\ref{canonical equation}).

Finally, we like to address that  the canonical momenta $\pi_i$ and
the Hamiltonian $H_{cR}$ are not the physically conserved  momentum
and the energy of the particle respectively, but they will play
important role in the quantization of $\mathcal{SR}_{cR}$-mechanics.

\subsection{Space-time symmetry of de Sitter special relativity and
the Neother charges}

\noindent The space time transformations {preserving} the Beltrami
metric were discovered about {30 years ago} by Lu, Zou and Guo
(LZG)\cite{look}\cite{Lu74}(see also Appendix of \cite{Yan1}).
   When we transform from one initial Beltrami frame $ x^{\mu}$ to
another Beltrami frame $ \tilde{x}^{\mu}$, and {when } the origin of
the new frame is $a^{\mu}$ in the original frame, {the}
transformations between them with 10 parameters  is as follows
% law,which first derived by
%Lu-qikeng. The transformation are checked in the Appendix.

\begin{eqnarray}\label{transformation}
x^{\mu} \;-\hskip-0.10in\longrightarrow\hskip-0.4in^{LZG}
 ~~ \tilde{x}^{\mu} &=& \pm \sigma(a)^{1/2} \sigma(a,x)^{-1}
(x^{\nu}-a^{\nu})D_{\nu}^{\mu}, \\
    \nonumber D_{\nu}^{\mu} &=& L_{\nu}^{\mu}+R^{-2} \eta_{\nu
\rho}a^{\rho} a^{\lambda} (\sigma
(a) +\sigma^{1/2}(a))^{-1} L_{\lambda}^{\mu} ,\\
\nonumber L : &=& (L_{\nu}^{\mu})\in SO(1,3), \\
\nonumber \sigma(x)&=& 1-{1 \over R^2}{\eta_{\mu \nu}x^{\mu} x^{\nu}}, \\
\nonumber \sigma(a,x)&=& 1-{1 \over R^2}{\eta_{\mu \nu}a^{\mu}
x^{\nu}}.
\end{eqnarray}
It will be called as LZG-transformation hereafter. Under
LZG-transformation, the $B_{\mu\nu}(x)$ and the action of
$\mathcal{SR}_{cR}$ transfer respectively as follows
\begin{equation} \label{B01}
 B_{\mu\nu}(x)\;-\hskip-0.10in\longrightarrow\hskip-0.4in^{LZG}
 ~~ ~\widetilde{B}_{\mu\nu}(\widetilde{x})={\pa x^\lambda \over \pa
 \widetilde{x}^\mu}{\pa x^\rho \over \pa
 \widetilde{x}^\nu}B_{\lambda\rho}(x)=B_{\mu\nu}(\widetilde{x}),
\end{equation}
\begin{equation} \label{B02}
 S_{cR}\equiv\int dt L_{cR}(t)=-m_0c\int dt {\sqrt{B_{\mu\nu}(x)dx^\mu dx^\nu} \over dt}
 \;-\hskip-0.05in-\hskip-0.05in\longrightarrow\hskip-0.4in^{LZG}
 ~ ~~\widetilde{S}_{cR}=S_{cR}.
\end{equation}
By the mechanics principle, this action {invariance indicates that}
there are 10 conserved Noether charges in $\mathcal{SR}_{cR}$ like
the $\mathcal{SR}_{c}$ case. For $\mathcal{SR}_{c}$ the Noether
charges are(e.g., see {\it pp581-586} and {\it Part 9} in
ref.\cite{Noether}):
\begin{eqnarray}\label{503a}
\begin{array}{rcl}
  &&{\rm{ Noether}\;charges\;for\;Lorentz\;boost:\;} ~~
 K_{c}^i=m_0 \gamma c (x^i- t \dot{x}^i) \\
 &&{\rm
Charges\;for\;space-transitions\;(momenta):}~~~  P_{c}^i=m_0 \gamma \dot{x}^i, \\
 &&{\rm Charge\;for\;time-transition\;(energy): }~~~
 E_{c}= m_0 c^2 \gamma \\
&&{\rm Charges\;for\;rotations\;in\;space\;(angular momenta):}~~~
L_{c}^i = \epsilon^{i}_{jk}x^{j}P_c^{k}.
\end{array}
\end{eqnarray}
Here $\gamma=\frac{1}{\sqrt{1-\frac{v^2}{c^2}}}$. Note the Noether
charges here are the same as the corresponding canonical quantities,
because the Lagrangian for $\mathcal{SR}_{c}$ is time-independent
and all the coordinates are cyclic. While in $\mathcal{SR}_{cR}$
there is no cyclic ignorable coordinates and the Lagrangian is
space-time dependent.

When  space rotations were neglected temporarily  for simplify, the
LZG-transformation both due to a Lorentz-like boost and a
space-transition in the $x^1$ direction with  parameters
$\beta=\dot{x}^1/c$ and $a^1$ respectively and due to a time
transition with  parameter $a^0$ can be explicitly written as
follows:
\begin{eqnarray}\label{general transformation}
\begin{array}{rcl}
t\rightarrow \tilde{t}&=& \frac{\sqrt{\sigma(a)}}{c \sigma(a,x)}
\gamma \left[ct-\beta x^1-a^0+ \beta a^1 +\frac{a^0-\beta
a^1}{R^2}\frac{a^0 ct-a^1 x^1-(a^0)^2 +(a^1)^2 }
{ \sigma(a)+\sqrt{\sigma(a)}} \right] \\
 x^1\rightarrow \tilde{x}^1&=& \frac{\sqrt{\sigma(a)}}{
\sigma(a,x)}\gamma \left[ x^1-\beta ct +\beta a^0 -a^1 +\frac{a^1-
\beta a^0}{R^2}
\frac{a^0 ct-a^1 x^1-(a^0)^2 +(a^1)^2}{ \sigma(a)+\sqrt{\sigma(a)}}\right]\\
 x^2\rightarrow
\tilde{x}^2&=&\frac{\sqrt{\sigma(a)}}{\sigma(a,x)}x^2 \\
 x^3\rightarrow
\tilde{x}^3&=&\frac{\sqrt{\sigma(a)}}{\sigma(a,x)}x^3
\end{array}
\end{eqnarray}
It is easy to check when $R\rightarrow \infty$ the above
transformation goes back to Poincar\'e transformation. Notice that
in the LZG-transformation there are 3 boost parameters
$\beta^{i}={\dot{x}^i \over c}=\frac{v^i}{c}$,  4 spacetime
transition parameters $(a^0,a^1,a^2,a^3)$ ( and 3 rotation
parameters $\theta^i$).  Here $(a^0,a^1,a^2,a^3)$ is the origin of
the resulting Beltrami initial frame
 in the original Beltrami frame. By the standard manner and
 eq.(\ref{general transformation}), we have got all $\mathcal{SR}_{cR}$-Noether
 charges in \cite{Yan1}, which
correspond to the $\mathcal{SR}_c$-Noether charges eq.(\ref{503a}).
Those $\mathcal{SR}_{cR}$-Noether charges are follows \cite{Yan1}
\begin{eqnarray}\label{503b}
\begin{array}{rcl}
  &&{\rm{ Noether}\;charges\;for\;Lorentz\;boost:\;} ~~
 K_{cR}^i=m_0 \Gamma c (x^i- t \dot{x}^i) \\
 &&{\rm
Charges\;for\;space-transitions\;(momenta):}~~~  P_{cR}^i=m_0 \Gamma \dot{x}^i, \\
 &&{\rm Charge\;for\;time-transition\;(energy): }~~~
 E_{cR}= m_0 c^2 \Gamma \\
&&{\rm Charges\;for\;rotations\;in\;space\;(angular momenta):}~~~
L_{cR}^i = \epsilon^{i}_{jk}x^{j}P_{cR}^{k},
\end{array}
\end{eqnarray}
where $\Gamma$ were given in eq(\ref{new parameter}). Compactly, by
the above, we have the 4-momentum in $\mathcal{SR}_{cR}$ as follows
\begin{eqnarray}\label{70-p4}
p_{cR}^\mu \equiv \{p_{cR}^0,~p_{cR}^i\} = m_0 \Gamma \dot{x}^\mu
={m_0c\over \sigma(x)}{dx^\mu \over ds}=-{1\over
\sigma(x)}B^{\mu\nu}\pi_\nu.
\end{eqnarray}

In terms of eq.(\ref{dispersion}), the Eienstein's famous
mass-energy-momentum formula $E_c^2=m_0^2c^4+c^2{\bf p}_c^2$ now
becomes
\begin{eqnarray} \label{sesan}
E_{cR}^2 =m_0^2 c^4+c^2{\mathbf p}_{cR}^2 + \frac{c^2}{R^2}
({\mathbf L}_{cR}^2-{\mathbf K}_{cR}^2),
\end{eqnarray}
where $E_{cR},{\mathbf p}_{cR},{\mathbf L}_{cR},{\mathbf K}_{cR}$
are conserved physical energy, momentum, angular-momentum and boost
charges in eq.(\ref{503b}) respectively.

\section{ quantum mechanics  in de Sitter special relativity}

\noindent Lagrangian-Hamiltonian formulation of mechanics is the
foundation of quantization. When the classical Poisson brackets in
canonical equations for canonical coordinates and canonical momentum
become operator's commutators, i.e., $\{x,\pi\}_{PB}\Rightarrow
{1\over i\hbar}[x,\hat{\pi}]$, the classical mechanics will be
quantized. In this way, for instance, the ordinary relativistic
(i.e., $\mathcal{SR}_c$) one-particle quantum equations have been
derived. To the particle with spin-0, that is just the well known
Klein-Gordon equation.

 Following this first principle clue, we have
derived the one-particle quantum mechanics  for $\mathcal{SR}_{cR}$
in \cite{Yan1}. In the canonic quantization formulism , the canonic
variable operators are $x^i, \; \hat{\pi}_i $ with $i=1,2,3 $. And
due to eq.(\ref{Poisson}) the basic commutators for the free
particle quantization theory of $\mathcal{SR}_{cR}$ are the same as
usual, i.e.,
\begin{eqnarray}\label{ij}
[x^i, \;{\pi}_j]=i\hbar \delta^i_j,~~~~~ [{\pi}_i, \;{\pi}_j]=0,
~~~~~[x_i, \;x_j]=0,
\end{eqnarray}
hereafter the hat notations for operators are removed. Considering
Wyle ordering of $(\pi x)$ and solving (\ref{ij}), we have
\cite{Yan1}
\begin{eqnarray}\label{quantum operator}
\pi_{\mu} =-i\hbar B^{-\frac{1}{4}}\pa_{\mu} B^{1\over 4} =-i\hbar
\pa_\mu -i\hbar B^{-\frac{1}{4}}(\pa_{\mu} B^{1\over 4}),
\end{eqnarray}
where $B=det(B_{\mu \nu})$. The classical dispersion relation
(\ref{dispersion}) can be rewritten as symmetric version $
B^{-\frac{1}{4}}\pi_{\mu}B^{\frac{1}{4}} B^{\mu\nu}
B^{\frac{1}{4}}\pi_{\nu}B^{-\frac{1}{4}}=m_{0}^2 c^2  $, and then
the $\mathcal{SR}_{cR}$-one particle wave equation reads
\begin{eqnarray}\label{KG}
B^{-\frac{1}{4}}\pi_{\mu}B^{\frac{1}{4}} B^{\mu\nu}
B^{\frac{1}{4}}\pi_{\nu}B^{-\frac{1}{4}}\phi(x,t)&=&m_{0}^2 c^2
\phi(x,t)\;,
\end{eqnarray}
where $\phi(x,t)$ is the particle's wave function.  Substituting
(\ref{quantum operator}) into (\ref{KG}), we have
\begin{equation}\label{KG in curved spactime}
\frac{1}{\sqrt{B}}\pa_{\mu}(B^{\mu\nu}\sqrt{B}\pa_{\nu})\phi+\frac{m_{0}^2
c^2}{\hbar^2} \phi=0,
\end{equation}
which is just the Klein-Gordon equation in  curved space-time with
Beltrami metric $B_{\mu\nu}$, and its explicit form is
\begin{equation}\label{KG1}
(\eta^{\mu\nu}-\frac{x^{\mu}x^{\nu}}{R^2})\partial_{\mu}\partial_{\nu}\phi
- 2\frac{x^{\mu}}{R^2}\partial_{\mu}\phi+\frac{m_{0}^2c^2}{\hbar^2
\sigma(x)} \phi = 0,
\end{equation}
which is the desired $\mathcal{SR}_{cR}$-quantum mechanics equation
for free particle with spin 0. Substituting (\ref{quantum operator})
into (\ref{70-p4}), we obtain the physical momentum and energy
operators (noting the subscripts $cR$ for
$p_{cR}^\mu,\;L_{cR}^{\mu\nu}$ in (\ref{70-p4}) will be moved
hereafter):
\begin{eqnarray}\label{momentum operator}
p^{\mu} &=& i\hbar
[(\eta^{\mu\nu}-\frac{x^{\mu}x^{\nu}}{R^2})\partial_{\nu}+\frac{5x^{\mu}}{2R^2}].
\end{eqnarray}
Operator $p^\mu$ together with  operator $ L^{\mu\nu} = (x^{\mu}
p^{\nu} -x^{\nu} p^{\mu})/(i\hbar)$ form a  algebra as follows
\begin{eqnarray} \label{ds algebra}
[p^{\mu},p^{\nu}]&=&\frac{\hbar^2}{R^2}L^{\mu\nu}
\\ \nonumber
[L^{\mu\nu},p^{\rho}]&=&\eta^{\nu\rho}p^{\mu}-\eta^{\mu\rho}p^{\nu}\\
\nonumber [L^{\mu\nu},L^{\rho\sigma}]&=&\eta^{\nu\rho}L^{\mu\sigma}-
\eta^{\nu\sigma}L^{\mu\rho}+\eta^{\mu\sigma}L^{\nu\rho}-\eta^{\mu\rho}L^{\nu\sigma}
\end{eqnarray}
which is just the de-Sitter algebra SO(1,4).  This fact means that
the quantization scheme presented in this paper preserves the
external space-time symmetry of $\mathcal{SR}_{cR}$.

By  the Klein-Gordon equation in  curved space-time with Beltrami
metric $B_{\mu\nu}$, eq.(\ref{KG in curved spactime}), we have the
corresponding Dirac equation which describes the particle with spin
$1/2$ \cite{Ut}\cite{NY}:
\begin{equation}\label{Cur-Dirac}
\left(ie_a^\mu \gamma^a D_\mu-{m_0c \over \hbar}\right)\psi=0,
\end{equation}
where $e_a^\mu$ is the tetrad and $D_\mu$ is the covariant
derivative with Lorentz spin connection $\omega^{ab}_\mu$. Their
definitions and relations are follows (e.g., see \cite{NY})
\begin{eqnarray}
\nonumber D_\mu &=&\pa_\mu-{i\over 4}\omega^{ab}_\mu\sigma_{ab},\\
\nonumber \{\gamma^a, \gamma^b\}&=& 2\eta^{ab},~~\sigma_{ab}={i\over
2}[\gamma_a, \gamma_b],~~{i\over 2}[\sigma_{ab}, \sigma_{cd}]=\eta_{ac}\sigma_{bd}
-\eta_{ad}\sigma_{bc}+\eta_{bd}\sigma_{ac}-\eta_{bc}\sigma_{ad}, \\
\nonumber e_\mu^a e_\nu^b \eta_{ab}&=&B_{\mu\nu},~~e_\mu^a e_\nu^b
B^{\mu\nu}=\eta^{ab},~~e^\mu_{a\; ;\nu}=\pa_\nu e_a^\mu+ \omega_{a\;\;\nu}^{\;\;b}e_b^\mu+\Gamma^\mu_{\lambda\nu} e^\lambda_a=0,\\
\nonumber \omega^{ab}_\mu &=& {1\over 2}(e^{a\rho}\pa_\mu e^b_\rho
-e^{b\rho}\pa_\mu e^a_\rho ) -{1\over
2}\Gamma^\rho_{\lambda\mu}(e^{a\lambda}e^b_\rho
-e^{b\lambda}e^a_\rho ),\\
\label{47} \Gamma^\rho_{\lambda\mu} &=& {1\over
2}B^{\rho\nu}(\pa_\lambda B_{\nu\mu} +\pa_\mu B_{\nu\lambda}
-\pa_\nu B_{\lambda\mu}).
\end{eqnarray}
It is straightforward  to check that the components
$\psi_\alpha\;(\alpha=1,\cdots 4)$ of the spinor satisfy the
Klein-Gordon equation (\ref{KG in curved spactime}).

\section{hydrogen atom in earth-QSO reference frame and Variation of
Electromagnetic Fine-Structure Constant}

\noindent Now we are going to solve $\mathcal{SR}_{cR}$-Dirac
equation for hydrogen atom on QSO. In this cosmologic quantum
system, there are two cosmologic length scales: cosmic radius $R\sim
10^{12}ly$ and the distance between QSO  and earth $c\;t$: $R>c\;t>
10^8 ly$, and two microcosmic length scales: the Compton wave length
of electron $a_c=\hbar/(m_e c)\simeq 0.3\times 10^{-12}m$, and Bohr
radius $a= \hbar^2/(m_e e^2)\simeq 0.5\times 10^{-10}m$. The
calculations for our purpose will be accurate up to
$\mathcal{O}(c^2t^2/R^2)$. The terms proportional to
$\mathcal{O}(c^4t^4/R^4)$, $\mathcal{O}(cta_c/R^2)$,
$\mathcal{O}(cta/R^2)$ etc will be omitted.

\subsection{$\mathcal{SR}_{cR}$-Dirac equation for hydrogen atom on QSO}
\noindent The phenomenology of atomic physics at the cosmologic
space-time scale should be discussed  in terms of
$\mathcal{SR}_{cR}$-quantum mechanics rather than
$\mathcal{SR}_{c}$'s. Now, we show the $\mathcal{SR}_{cR}$-Dirac
equation of hydrogen atom on a QSO in the earth-QSO reference frame.
As illustrated in Fig.1, the earth locates at the origin of frame,
the proton (nucleus of hydrogen atom) locates at $Q=\{Q^0\equiv
c\;t,\;Q^1=c\;t,\;Q^2=0,\;Q^3=0\}$, which is on QSO-light-cone
$B_{\mu\nu}(Q)Q^\mu Q^\nu=\eta_{\mu\nu}Q^\mu Q^\nu=0$. The metric of
the space-time near $Q$ is
\begin{equation}\label{metric5}
B_{\mu\nu}(Q)=\eta_{\mu\nu}+{1\over R^2}\eta_{\mu \lambda
}Q^\lambda\eta_{\nu \rho}Q^\rho,\;{\rm and\;hence}~~
B_{ij}(Q)=\eta_{ij}+{c^2 t^2\over R^2} \delta_{i1}\delta_{j1}.
\end{equation}
 The electron's coordinates  are $L=\{L^0\equiv
c\;t,\;L^1,\;L^2,\;L^3\}$, and the relative space coordinates
between proton and electron are $x^i=L^i-Q^i$.
\begin{figure}[hptb]
\begin{center}
\includegraphics[width=0.56\textwidth]{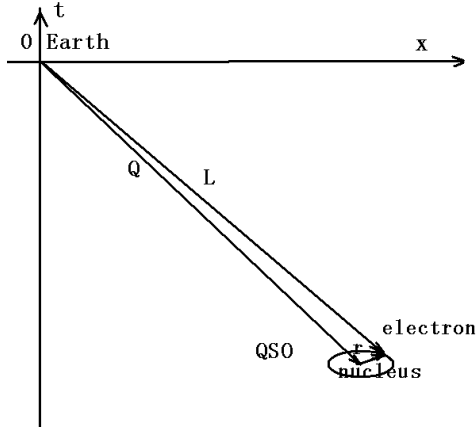}
\vskip-2.5in \caption{\label{Fig1} \small Sketch of the earth-QSO
reference frame. The earth locates at the origin. The position
vector for nucleus of atom on QSO is $Q$, and for electron is $L$.
The distance between nucleus and electron is $r$. }
\end{center}
\end{figure}
The magnitude of $r\equiv \sqrt{-\eta_{ij}x^ix^j}\sim a$ (where
$a\simeq 0.5\times 10^{-10}m$ is Bohr radius), and $|x^i|\sim a$.

 According to gauge principle, the
electrodynamic interaction between the nucleus and the electron can
be taken into account by replacing the operator $D_\mu$ in
eq.(\ref{Cur-Dirac}) with the
 $U(1)$-gauge covariant derivative $\mathcal{D}_\mu^L \equiv
D^L_\mu-\delta_{\mu 0}ie/(c\hbar)\phi(x)$. Hence, the
$\mathcal{SR}_{cR}$-Dirac equation for electron in hydrogen at QSO
reads
\begin{equation}\label{Dirac1}
(ie^\mu_a \gamma^a\mathcal{D}_\mu^L -{\mu c\over \hbar })\psi=0,
\end{equation}
where $\mu=m_e/(1+{m_e\over m_p})$ is the reduced mass of electron,
$\mathcal{D}_\mu^L={\pa\over \pa L^\mu}-{i\over
4}\omega^{ab}_\mu\sigma_{ab}-\delta_{\mu 0}ie/(c\hbar)\phi(x),$
$e^\mu_a$  and $\omega_\mu^{ab}$ have been given in
eqs.(\ref{star28}) (\ref{47}). For our purpose, we approximate
$e^\mu_a$ and $\omega^{ab}_\mu$ up to $\mathcal{O}(1/R^2)$,
\begin{eqnarray}\label{e}
e^\mu_a &=&\left(1-{\eta_{cd}L^cL^d\over
2R^2}\right)\eta^\mu_a-{\eta_{ab}L^bL^\mu \over
2R^2}+\mathcal{O}(1/R^4),\\
\label{omega} \omega^{ab}_\mu &=& {1\over 2R^2}(\eta^a_\mu L^b-
\eta^b_\mu L^a )+\mathcal{O}(1/R^4).
\end{eqnarray}

\subsection{Solution of usual $\mathcal{SR}_c$-Dirac
equation for hydrogen atom at QSO }

\noindent At first, we show the solution of usual
$\mathcal{SR}_c$-Dirac equation in the earth-QSO reference frame of
Fig.1, which serves as leading order of solution for the
$\mathcal{SR}_{cR}$-Dirac equation with $R\rightarrow \infty$ in
that reference frame. For the hydrogen, $\pa_\mu\rightarrow
\mathcal{D}^L_\mu=\pa^L_\mu-\delta_{\mu 0}ie/(c\hbar )\phi_M(x)$
(noting $\omega_{\mu}^{ab}|_{R\rightarrow \infty}=0$), where
$\phi_M(x)$ is nucleus electric potential at $x^i$ in Minkowski
space defined by following equation
\begin{equation}\label{potential}
-\eta^{ij}\pa_i\pa_j\phi_M(x)=\nabla^2\phi_M(x)=-4\pi\rho(x)=-4\pi
e\delta^{(3)}(\mathbf{x}).
\end{equation}
The solution is $\phi_M(x)=e/r$, and hence $\pa_0\rightarrow
\mathcal{D}^L_0=\pa_0-ie^2/(c\hbar r).$ Then, the
$\mathcal{SR}_c$-Dirac equation reads
\begin{equation}\label{Dirac2}
i\hbar \pa_t\psi=\left(-i\hbar c \vec{\alpha} \cdot \nabla_{L}+\mu
c^2\beta -{e^2\over r}\right)\psi,
\end{equation}
where $\beta=\gamma^0,\; \alpha^i=\beta\gamma^i$. Noting the nucleus
position $\mathbf{Q}=$constant, we have
\begin{equation}\label{D3}
\nabla_{L}={\pa \over \pa \mathbf{L}} = {\pa \over \pa
(\mathbf{Q}+\mathbf{r})}={\pa \over \pa \mathbf{r}}\equiv\nabla  ,
\end{equation}
and eq.(\ref{Dirac2}) becomes the standard Dirac equation for
electron in hydrogen at its nucleus reference frame. Energy $E$ for
$\mathcal{SR}_c$-mechanics is conserved, and the hydrogen is the
stationary states of $\mathcal{SR}_c$-Dirac equation. The stationary
state condition is
\begin{equation}\label{S1}
i\hbar \pa_t\psi=E\psi.
\end{equation}
As is well known, combining eqs.(\ref{Dirac2}), (\ref{D3}) with
(\ref{S1}), we have
\begin{equation}\label{Dirac4}
E\psi=\left(-i\hbar c \vec{\alpha} \cdot \nabla +\mu c^2\beta
-{e^2\over r}\right)\psi,
\end{equation}
which is the stationary $\mathcal{SR}_{c}$-Dirac equation for
hydrogen. The problem has been solved in terms of standard way, and
the results are follows (see, e.g.,
\cite{shiff}\cite{Messian}\cite{zeng})
\begin{eqnarray}\label{solution1}
E=E_{n,K}&=&\mu c^2\left[1+{\alpha^2 \over
(\sqrt{K^2-\alpha^2}+n_r)^2} \right]^{-1/2} \\ \nonumber
&&\alpha\equiv {e^2\over \hbar c},~~~~n_r=0,1,\;2,\;\cdots \\
\nonumber &&|K|=(j+1/2)=1,\; 2,\; 3,\; \cdots.
\end{eqnarray}
And its expansion equation in $\alpha$ is
\begin{equation}\label{usual}
E=\mu c^2-\mu c^2{\alpha^2\over 2n^2}\left[1+{\alpha^2\over
n^2}\left( {n\over j+1/2}-{3\over 4}\right)+\cdots\right],\;\;\;
n=n_r+|K|=1,2,3\cdots.
\end{equation}
The corresponding hydrogen's wave functions $\psi$ have  also
already been finely derived (see e.g.,
\cite{shiff}\cite{Messian}\cite{zeng}). The complete set of
commutative observables is $\{H,\; K,\;\mathbf{j}^2,\;j_z\}$, so
that $\psi=\psi_{n, K, j, j_z}(\mathbf{r},\hbar, \mu, \alpha) $,
where $\mathbf{j}=\mathbf{l}+{\hbar\over 2} \mathbf{\Sigma},\;\hbar
K=\beta(\mathbf{\Sigma}\cdot \mathbf{l}+\hbar)$, and $\alpha=e^2/
(\hbar c)$.

\subsection{Beltrami-geometry effects  in  $\mathcal{SR}_{cR}$-Dirac
equation } \noindent By eqs.(\ref{Dirac1}), (\ref{e}) and
$\pa_\mu\rightarrow \mathcal{D}_\mu^L={\pa\over \pa L^\mu}-{i\over
4}\omega^{ab}_\mu\sigma_{ab}-\delta_{\mu 0}ie/(c\hbar)\phi_B(x)$,
(where $\phi_B(x)$ is the electric potential in Beltrami space), we
have the
 $\mathcal{SR}_{cR}$-Dirac equation for the electron in hydrogen at
 the earth-QSO reference frame as follows
\begin{equation}\label{Dirac5}
\hbar c \beta \left[i\left( 1-{\eta_{ab}L^aL^b\over
2R^2}\right)\gamma^\mu \mathcal{D}_\mu^L -{i\over 2R^2} \eta_{ab}
L^a\gamma^b L^\mu \mathcal{D}_\mu^L-{\mu c\over \hbar}
\right]\psi=0,
\end{equation}
where factor $\hbar c \beta$ in the front of the equation is only
for convenience. We expand each terms of (\ref{Dirac5}) in order as
follows:
\begin{enumerate}
\item Since observed QSO must locates at the light cone, then
$\eta_{ab}L^aL^b\simeq 0$, and the first term of (\ref{Dirac5})
reads
\begin{equation}\label{first}
\hbar c \beta i \gamma^\mu \mathcal{D}^L_\mu \psi=\left(i\hbar \pa_t
+i\hbar c \vec{\alpha}\cdot \nabla + {\hbar c \beta \over 4}
\omega^{ab}_\mu\gamma^\mu\sigma_{ab} +e\phi_B(x)\right) \psi,
\end{equation}
where, being similar to (\ref{potential}), $\phi_B(x)$ defined by
following equation
\begin{equation}\label{potential-B}
-B^{ij}(Q)\pa_i\pa_j\phi_B(x)\hskip-0.07in=\hskip-0.06in\left(\nabla^2+{c^2t^2\over
R^2}{\pa^2\over \pa (x^1)^2}\right)\hskip-0.06in
\phi_B(x)\hskip-0.06in=\hskip-0.06in-4\pi\rho_B(x)\hskip-0.06in=\hskip-0.06in{-4\pi
e\over \sqrt{-det(B_{ij}(Q))}}\delta^{(3)}(\mathbf{x}).
\end{equation}
The solution is (see Appendix A)
\begin{equation}\label{potential-B1}
\phi_B={e\over r_B}\simeq {e\over r}\left(1+{c^2t^2(x^1)^2\over
2R^2r^2} \right),
\end{equation}
where $r_B=\sqrt{(\td{x}^1)^2+(x^2)^2+(x^3)^2}$ with
$\td{x}^1=(1-c^2t^2/(2R^2))x^1$. The correction factor due to
$B^{ij}$ shows a little bit of non-isotropy in
$\hat{x}^1$-direction. In order to deal with this non-isotropy
effect, we will use $\{\td{x}^1,\;x^2,\;x^3\}$ (instead of usual
$\{x^1,\;x^2,\;x^3\}$) to be the space coordinate variables of Dirac
equation  \footnote{ The author thanks Professor Chao-Guang Huang
for this suggestion.}. Following notations are introduced hereafter:
\begin{eqnarray}\label{nota1}
\mathbf{r}_B&=&\mathbf{i}\td{x}^1+ \mathbf{j}x^2+\mathbf{k}x^3,~~~~|\mathbf{r}_B|=r_B, \\
\label{nota2} \nabla_B &=& \mathbf{i}{\pa \over \pa\td{x}^1}+
\mathbf{j}{\pa \over \pa x^2}+\mathbf{k}{\pa\over \pa
x^3},~~~~~\td{x}^i\in\{\td{x}^1,\;x^2,\;x^3\}.
\end{eqnarray}
Then the eq.(\ref{first}) becomes
\begin{equation}\label{first1}
\hbar c \beta i \gamma^\mu \mathcal{D}^L_\mu \psi=\left(i\hbar \pa_t
+i\hbar c \vec{\alpha}\cdot \nabla_B-i\hbar c{c^2t^2\over
2R^2}\alpha^1{\pa\over \pa\td{x}^1} + {\hbar c \beta \over 4}
\omega^{ab}_\mu\gamma^\mu\sigma_{ab} +{e^2\over r_B} \right) \psi.
\end{equation}

\item Estimating the contributions of the fourth term in RSH of (\ref{first1})
( the spin-connection contributions): By (\ref{omega}), the ratio of
the fourth term to the first term of (\ref{first1}) is:
\begin{equation}\label{61}
\left|{ {\hbar c \beta \over 4}
\omega^{ab}_\mu\gamma^\mu\sigma_{ab}\psi \over i\hbar \pa_t\psi }
\right|\sim {\hbar c\over 4}{ct\over 2R^2}{1\over m_ec^2}={ct\over
8R^2}{\hbar\over m_ec}={1\over 8}{cta_c\over R^2}\sim 0,
\end{equation}
where $a_c= \hbar/(m_e c)\simeq 0.3\times 10^{-12}m$ is the Compton
wave length of electron. $\mathcal{O}(cta_c/R^2)$-term is
neglectable. Therefore the 3-rd term in RSH of (\ref{first}) has no
contribution to our approximation calculations.

\item Substituting  (\ref{61}) into
(\ref{first1}) and noting $\eta_{ab}L^aL^b\simeq 0$, we get the
first term in LHS of (\ref{Dirac5})
\begin{equation}\label{first-1}
\hbar c \beta i\left( 1-{\eta_{ab}L^aL^b\over 2R^2}\right)
\gamma^\mu \mathcal{D}^L_\mu \psi=\left(i\hbar \pa_t +i\hbar c
\vec{\alpha}\cdot \nabla_B-i\hbar c{c^2t^2\over
2R^2}\alpha^1{\pa\over \pa\td{x}^1}  +{e^2\over r_B} \right) \psi.
\end{equation}

\item The second term of (\ref{Dirac5}) is
\begin{eqnarray}\label{second}
 \nonumber -\hbar c \beta {i\over 2R^2} \eta_{ab} L^a\gamma^b L^\mu
\mathcal{D}^L_\mu \psi&=&-{i\hbar c \beta \over 2R^2}(\gamma^0
L^0-\vec{\gamma}\cdot \vec{L})L^\mu \left[\pa_\mu^L-\delta_{\mu
0}{ie^2\over c\hbar r_B}\right]\psi +\mathcal{O}({1\over R^4})
\\ \nonumber &=&-{i\hbar c \over
2R^2}(L^0-\vec{L}\cdot\vec{\alpha})\left( L^0\pa_0^L-L^0{ie^2\over
c\hbar r_B}+L^i\pa_i^L\right)\psi \\
\nonumber &\simeq& -{ic\hbar \over
2R^2}(L^0-L^1\alpha^1)(L^0\pa_0^L-L^0{ie^2\over
c\hbar r_B}+L^1\pa^L_1)\psi \\
\nonumber &\simeq& -{ic\hbar \over 2R^2}[\left(1-{L^1\over
L^0}\alpha^1\right)(L^0)^2\left(\pa_0^L-{ie^2\over
c\hbar r_B}\right)\\
&&+L^0L^i\pa^L_i-L_1^2\alpha_1\pa_1^L]\psi,
\end{eqnarray}
where following estimations are used
\begin{equation}\label{note1}
{L^2\over R}\sim {L^3\over R}\sim {a\over R}\sim 0.
\end{equation}
In order to simplify (\ref{second}) further, we note that
$c\vec{\alpha}=\mathbf{v}$ in Dirac equation theory, and hence
\begin{eqnarray}\label{note2}
{1\over R^2}{L^1\over L^0}\alpha_1\psi&\simeq& {1\over R^2}{v^1\over
c}<< {1\over R^2}\\
\label{note3} \nonumber {1\over R^2}L^i\pa_i^L\psi &=&{1\over
R^2}(Q^i+\td{x}^i){\pa\over\pa\td{x}^i}\psi+\mathcal{O}(1/R^4)\\
\nonumber &\Rightarrow& {i\over \hbar R^2}[\langle Q^ip_i\rangle
+\langle \td{x}^ip_i\rangle] ={i\over \hbar R^2}[\langle
\vec{Q}\cdot\vec{p}\rangle +\langle
\vec{r}_B\cdot\vec{p}\rangle]\\
\label{note3}&=&0.
\end{eqnarray}
where $Q^i{\pa\over \pa\td{x}^i}\psi\Rightarrow \langle
\vec{Q}\cdot\vec{p}\rangle$ means that $\langle
\vec{Q}\cdot\vec{p}\rangle$ serves as mean-value of operator $
\vec{Q}\cdot\vec{p}$ and can be the leading order of the operator's
approximate expansion. Sine the electron does  circular motion
around the nucleus, and  is always inside atom, we have $\langle
\vec{Q}\cdot\vec{p}\rangle=\langle \vec{r}_B\cdot\vec{p}\rangle=0$,
and hence (\ref{note3}) holds.

Inserting (\ref{note2}) (\ref{note3}) into (\ref{second}), we have
\begin{eqnarray}\label{second1}
 \nonumber -\hbar c \beta {i\over 2R^2} \eta_{ab} L^a\gamma^b L^\mu
\mathcal{D}^L_\mu \psi&=&-{i\hbar c \beta \over 2R^2}(\gamma^0
L^0-\vec{\gamma}\cdot \vec{L})L^\mu \left[\pa_\mu^L-\delta_{\mu
0}{ie^2\over c\hbar r_B}\right]\psi +\mathcal{O}({1\over R^4})\\
\nonumber &\simeq& -{ic\hbar \over
2R^2}[(L^0)^2\left(\pa_0^L-{ie^2\over c\hbar
r_B}\right)-L_1^2\alpha_1\pa_1^L]\psi\\
&=&-i\hbar{c^2t^2\over 2R^2}\left(\pa_t-{ie^2\over \hbar
r_B}\right)\psi+i\hbar c{c^2t^2\over 2R^2}{\pa\over
\pa\td{x}^1}\psi.
\end{eqnarray}

\item Therefore, substituting (\ref{first-1}) (\ref{second1}) into by (\ref{Dirac5}), we
have
\begin{equation}\label{Dirac6}
i\hbar\left(1-{c^2t^2\over 2R^2}\right)\pa_t\psi=\left[ -i\hbar c
\vec{\alpha}\cdot\nabla_B + \mu c^2\beta -\left(1-{c^2t^2\over
2R^2}\right){e^2\over r_B}\right]\psi.
\end{equation}
This is a time-dependent wave equation.
 It is somehow difficult to deal with the time-dependent
 problems in quantum mechanics. Generally, there are two
approximative approaches to discuss two extreme cases respectively:
(i) The modification in states obtained by the wave equation depends
critically on the time $T$ during which the modification of the
system's "Hamiltonian" take place. For this case, one would use the
sudden approach; And, (ii), for case that of a very slow
modification of Hamiltonian,  the adiabatic approach works
\cite{Messian}. To wave equation of (\ref{Dirac6}), like the
discussions in Introduction of this paper, since $R$ is
cosmologically large and $R>>ct$, factor $ (c^2t^2/ R^2)$ makes the
time-evolution of the system is so slow that the adiabatic
approximation \cite{Born} may legitimately
 works. In the below (the subsection {\bf E}), we will provide a calculations to
 confirm this point.

\end{enumerate}

\subsection{ $\mathcal{SR}_{cR}$-Dirac
equation for  spectra of hydrogen}

\noindent  In order to discuss the spectra of hydrogen by
$\mathcal{SR}_{cR}$-Dirac equation, we need to find out its
solutions with certain physics energy $E$. By eq.(\ref{momentum
operator}), and being similar to
 (\ref{S1}), the $\mathcal{SR}_{cR}$-energy eigen-state condition for
 (\ref{Dirac6}) can be derived by means of the operator expression of momentum
 in $\mathcal{SR}_{cR}$ (\ref{momentum operator}):
\begin{eqnarray}\label{S2}
\nonumber p^0={E\over c}&=&i\hbar\left[{1\over c}\pa_t-{ct\over
R^2}x^\nu\pa^L_\nu+{5ct\over 2R^2}\right]\\
\nonumber E&=&i\hbar \left[\pa_t-{c^2t^2\over
R^2}\pa_t+{5ct\over 2R^2}\right]\\
E\psi&\simeq &i\hbar\left(1-{c^2t^2\over R^2}\right)\pa_t\psi,
\end{eqnarray}
where a estimation for the ratio of the 3-rd term to the 2-nd of
$E\psi$ were used:
$${|i\hbar {5c^2t\over 2R^2}\psi|\over |{-c^2t^2\over R^2}i\hbar
\pa_t\psi|}\sim {|i\hbar {5c^2t\over 2R^2}|\over |{-2c^2t^2\over
R^2}E|}\sim {5\hbar \over 2t m_e c^2}\equiv {5\over 2}{a_c\over ct}
$$  where $a_c\simeq 0.3\times 10^{-12}$m is the Compton wave length of electron and $ct$ is about the
distance between earth and QSO. In our approximative calculations
$a_c/(ct)$ is neglectable. For instance, to a QSO with $ct\sim
10^9$ly, $a_c/(ct)\sim 10^{-38}<< (ct)^2/R^2\sim 10^{-5}$. Hence the
3-rd term of $E\psi$ were ignored.

Inserting (\ref{S2}) into (\ref{Dirac6}), we have
$$\left(1+{c^2t^2\over 2R^2}\right)E\psi=
\left[ -i\hbar c \vec{\alpha}\cdot\nabla_B + \mu c^2\beta
-\left(1-{c^2t^2\over 2R^2}\right){e^2\over r_B}\right]\psi,$$ or
\begin{eqnarray}\label{Dirac7}
 E\psi=\left[ -i\hbar c\hskip-0.05in\left(1-{c^2t^2\over 2R^2}\right)
\vec{\alpha}\cdot\nabla_B
\hskip-0.06in+\hskip-0.06in\left(1-{c^2t^2\over 2R^2}\right) \mu
c^2\beta -\left(1-{c^2t^2\over R^2}\right){{e}^2\over r_B}
\right]\psi.
\end{eqnarray}
Eq.(\ref{Dirac7}) is the  $\mathcal{SR}_{cR}$-Dirac equation for
hydrogen spectra up to $\mathcal{O}(c^2t^2/R^2)$ (say again,
$\mathcal{O}(1/R^4),\;\mathcal{O}(cta/R^2),\;\mathcal{O}(cta_c/R^2)$
terms have been neglected). Eq. (\ref{Dirac7}) can further be
written as follows
\begin{eqnarray}\label{Dirac7-1}
 E\psi=\left( -i\hbar_t c
\vec{\alpha}\cdot\nabla_B + \mu_t c^2\beta -{{e}_t^2\over r_B}
\right)\psi,
\end{eqnarray}
where
\begin{eqnarray}\label{hme1}
\hbar_t&=& \left(1-{c^2t^2\over 2R^2}\right)\hbar, \\
\label{hme2}\mu_t&=& \left(1-{c^2t^2\over 2R^2}\right)\mu, \\
\label{hme3} e_t &=& \left(1-{c^2t^2\over 2R^2}\right)e.
\end{eqnarray}
Eq. (\ref{Dirac7-1}) is same as (\ref{Dirac4}) except
$\hbar,\;\mu,\;e$ be replaced by $\hbar_t,\;\mu_t,\;e_t$. However,
since the time $t$ is dynamic variable in the time-dependent
Hamiltonian system, we  do not know up to now whether $t$ can be
approximately treated as a parameter in the system. Hence, at this
stage we still cannot conclude $\hbar,\;\mu,\;e$ are time variations
by (\ref{hme1}), (\ref{hme2}), (\ref{hme3}). In the following, we
pursue this subject.

\subsection{Adiabatic approximation solution to $\mathcal{SR}_{cR}$-Dirac
spectra equation }

\noindent Comparing (\ref{Dirac7}) with (\ref{Dirac4}), we can see
that there are three correction terms in (\ref{Dirac7}), which are
proportional to $ (c^2t^2/ R^2)$. Those corrections service of the
effects of $\mathcal{SR}_{cR}$. In order to examine adiabatic
approach, we rewrite spectra equation (\ref{Dirac7}) into version of
wave equation like eq.(\ref{Dirac2}) via $E\Rightarrow i\hbar\pa_t$:
\begin{eqnarray}\label{Dirac8}
 i\hbar\pa_t\psi&= &H(t)\psi=[H_0(r, {e})+H'(t)]\psi, \\
\label{Dirac9} {\rm where}~~~~~~~~~H_0(r,{e})&=& -i\hbar c
\vec{\alpha}\cdot\nabla_B + \mu c^2\beta
-{{e}^2\over r_B}~ ~(see\; eq.(\ref{Dirac4}))\\
\label{Dirac10} H'(t)&=&-\left({c^2t^2\over 2R^2}\right) H_0(r,
{\sqrt{2}}{e}).
\end{eqnarray}
Suppose  initial state of the atom is
$\psi(t=0)=\psi_s(\mathbf{r}_B,\hbar,\mu,{\alpha})$ where
$s=\{n_s,\;K_s,\;\mathbf{j}_s^2,\;j_{sz}\}$, by eqs. (\ref{Dirac8})
(\ref{Dirac9}) (\ref{Dirac10}), and catching the time-evolution
effects, we have (see Chapter XVII of Vol II of \cite{Messian}, and
Appendix B)
\begin{equation}\label{wave1}
\psi(t)\simeq\psi_s(\mathbf{r}_B,\hbar_t,
\mu_t,\alpha_t)e^{-i{E_s\over \hbar}t}+\sum_{m\neq
s}{\dot{H}'(t)_{ms} \over i\hbar
\omega_{ms}^2}\left(e^{i\omega_{ms}t}-1\right)\psi_m(\mathbf{r}_B,\hbar_t,
\mu_t,\alpha_t)e^{\left(-i\int_0^t{E_m(\theta)\over
\hbar}d\theta\right)},
\end{equation}
where $\hbar_t$, $\mu_t$ are given in (\ref{hme1}) and (\ref{hme2}),
and
\begin{eqnarray}
\label{alpha} \alpha_t &\equiv& {e_t^2\over \hbar_t
c}=\left(1-{c^2t^2\over
2R^2}\right){\alpha},~~{\rm with}~~{\alpha}={{e}^2\over \hbar c}\\
\nonumber  \dot{H}'(t)_{ms}|_{(m\neq s)}&=&\langle m |\dot{H}'(t)|
s\rangle |_{(m\neq s)}={-c^2t\over R^2}\langle m |H_0(r,
{\sqrt{5\over 3}}{e})|
s\rangle |_{(m\neq s)}\\
\nonumber &=&{-c^2t\over R^2}\langle m |\left(H_0(r, {\sqrt{5\over
3}}{e})-H_0(r,e)\right)|
s\rangle |_{(m\neq s)}\\
 \nonumber &=&{-c^2t\over R^2}\langle
n_m,\;K_s,\;\mathbf{j}^2_s,\;j_{sz}|(-2+1)({{e}^2\over r})|
n_s,\;K_s,\;\mathbf{j}^2_s,\;j_{sz} \rangle
e^{-i(\omega_s-\omega_m)t} \\
\label{Hms}&=&{c^2t\over R^2}\langle n_m|{{e}^2\over r}| n_s \rangle
e^{-i(\omega_s-\omega_m)t},\\
\label{o} \omega_{ms}&=&\omega_m-\omega_s,~~\omega_m={E_m\over
\hbar}.
\end{eqnarray}
Note, formula $\langle m| H_0(r,e)|s\rangle |_{m\neq s}=0$ has been
used in the calculations of (\ref{Hms}). The second term of
Right-Hand-Side (RHS) of eq. (\ref{wave1}) represents the quantum
transition amplitudes from $\psi_s$-state to $\psi_m$, which belong
to non-adiabatic effect corrections (or the perturbation corrections
from the adiabatic approximation). Now for showing the order of
magnitude of such corrections, we estimate $|\dot{H}'(t)_{ms}/\hbar
\omega_{ms}^2|$ for $s=1,\;m=2$ and $t\sim 10^9{\rm Yr},\;R\sim
10^{12} {\rm ly}$. To the leading order of $\alpha$, the radial wave
functions for hydrogen with $n=1,\;2$ are
$$R_{10}={2\over a^{3/2}}\exp[-r/a],~~~~R_{20}={1\over
\sqrt{2}a^{3/2}}\left(1-{r\over 2a}\right)\exp[-r/2a],$$ where Bohr
radius $ a\simeq 0.5\times 10^{-10}m$. Therefore, we have
\begin{equation}\label{est}
\left|\dot{H}'(t)_{21}\over \hbar \omega_{21}^2\right|\simeq
{256\sqrt{2} \over 243}{ct \over  R^2}{a\over \alpha}\simeq
1.4\times 10^{-40}<<\left({c^2t^2\over R^2}\sim
\mathcal{O}(10^{-5})\right)<<1.
\end{equation}
Generally, the $|\dot{H}'(t)_{ms}/\hbar \omega_{ms}^2|_{m\neq s}\sim
|\dot{H}'(t)_{21}/\hbar \omega_{21}^2|$ are also tiny. The basic
reason is that the variation of the potential $H'(t)$ in the the
Bohr time-period $(\Delta
T_{ms}^{(Bohr)})\dot{H}'(t)_{ms}=(2\pi/\omega_{ms})\dot{H}'(t)_{ms}$
are much much less than $\hbar \omega_{ms}$. That  makes the quantum
transition from lower state $|s\rangle$ to higher state $|m \rangle$
almost impossible. Thus we conclude that the non-adiabatic effect
corrections are tiny, and by eq. (\ref{wave1}) the adiabatic wave
function of leading order is legitimate and  accurate enough as the
solution of eq. (\ref{Dirac8}):
\begin{equation}\label{wave2}
\psi(t)\simeq\psi_s(\mathbf{r}_B,\hbar_t,
\mu_t,\alpha_t)e^{-i{E_s\over \hbar}t}.
\end{equation}
Therefore the solution of (\ref{Dirac7}) is
$\psi=\psi_s(\mathbf{r}_B,\hbar_t, \mu_t,\alpha_t)$ with
$s=\{n,\;K,\;\mathbf{j}^2,\;j_{z}\}$ and
$\hbar_t,\;\mu_t,\;\alpha_t$ defined by (\ref{hme1}) (\ref{hme2})
(\ref{alpha}). For the solution of $\mathcal{SR}_{cR}$-Dirac
equation of hydrogen at earth-QSO reference framework, very
interesting result  is that the electromagnetic fine-structure
constant and the mass of electron are of variation with time as
follows (i.e., eqs. ((\ref{alpha}), (\ref{hme2}))
\begin{eqnarray}
\label{alpha3}{\Delta \alpha \over \alpha}&\equiv & {\alpha_t
-\alpha \over \alpha} = -{c^2t^2\over 2R^2},\\
\label{alpha2} {\Delta m_e \over m_e}&\equiv & { (m_{e})_t-m_e \over
m_e}= -{c^2t^2\over 2R^2}.
\end{eqnarray}
Because $ct$ represents  the distance between earth and QSO, above
equations  indicate that $\Delta\alpha/\alpha$ and $\Delta m_e/m_e$
can also be thought of variation with distances. The observation
quantity in experiments
\cite{webb06,Murphy03b,Murphy03a,murphy01b,murphy01a,webb01,webb99}
is the frequency of spectra $\omega_t$ that is as follows
\begin{eqnarray}\nonumber
\omega_t=E_t/\hbar_t&=&{\mu_t\over \hbar_t} c^2\left[1+{\alpha_t^2
\over (\sqrt{K^2-\alpha_t^2}+n_r)^2} \right]^{-1/2} \\
\label{solution2} &=&{\mu\over \hbar} c^2\left[1+{\alpha_t^2 \over
(\sqrt{K^2-\alpha_t^2}+n_r)^2} \right]^{-1/2},
\end{eqnarray}
where fact of $\mu_t/\hbar_t=\mu/\hbar$ due to (\ref{hme1})
 (\ref{hme2}) has been used. Consequently, the $t$-dependence of
 $\omega_t$ is caused by $t$-dependence of $\alpha_t$ totally, and
 hence the time variation of $\alpha$ could be observed by analyzing
the spectra emitted from atoms on distant galaxy.

Bing equivalent with (\ref{alpha2}), (\ref{alpha3}) and
(\ref{hme1}), and noting Compton wave length of electron
$a_c=\hbar/(m_ec)$ and Bohr radius $a=\hbar^2/(m_ee^2)=a_c/\alpha$,
we can also express the variations as follows
\begin{eqnarray}\label{Compton}
{\Delta a_c \over a_c}&\equiv& {(a_c)_t-a_c\over a_c}=0, \\
\label{Bohr}{\Delta a \over a}&\equiv& {a_t-a\over a}={c^2t^2\over
2R^2}.
\end{eqnarray}

\subsection{Comparing theory predictions to observations}

The observations of absorption spectra of distant interstellar
clouds were reported in
\cite{webb06,Murphy03b,Murphy03a,murphy01b,murphy01a,webb01,webb99}.
They belong to directly exploring cosmic atom physics
experimentally. Murphy and collaborators \cite{Murphy03b} studied
the spectra of 143 quasar absorption systems over the redshift range
$0.2< z_{abs}<4.2$. Their most robust estimate is a weighted mean
\begin{equation} \label{ex1}
{\Delta \alpha \over \alpha}=(-0.57\pm 0.11)\times 10^{-5}.
\end{equation}
Comparing  with the prediction (\ref{alpha3}), we conclude that
$R^2>0$. This means that the space-time symmetry for
$\mathcal{SR}_{cR}$ is de Sitter-$SO(4,1)$ instead of anti-de
Sitter-$SO(3,2)$.

\begin{figure}[ht]
\includegraphics[scale=1.0]{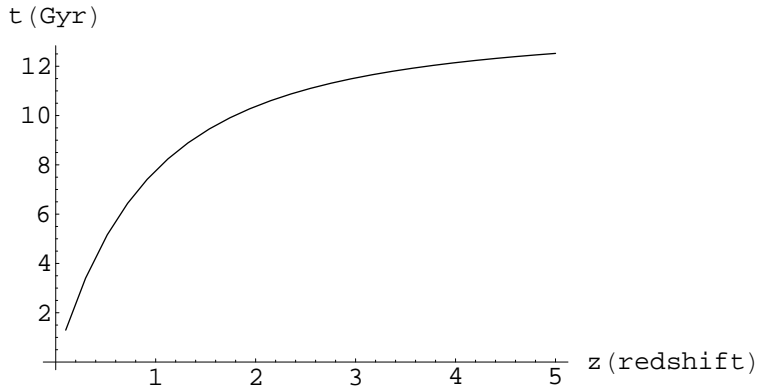}
\caption{\label{fig2}The $t-z$ relation in $\Lambda$CDM model
(eq.(\ref{La1})).}
\end{figure}

\begin{figure}[ht]
\includegraphics[scale=1.0]{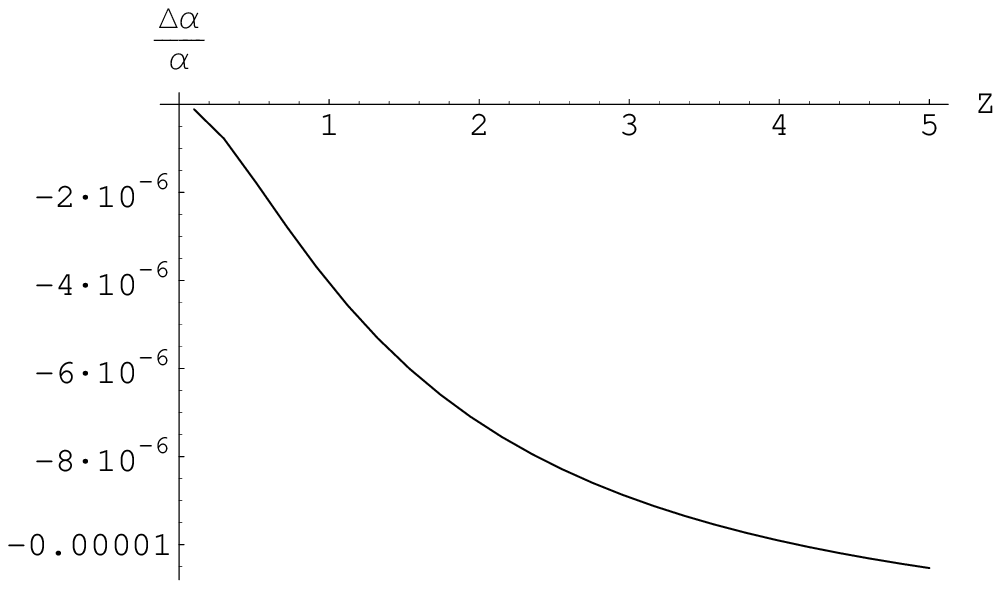}
\caption{\label{fig3}The $\Delta \alpha/\alpha$ as function of the
red shift $z$.}
\end{figure}

The 134 data points are assigned three epochs in ref. \cite{Dent}
(see table II), and the redshift $z$-dependence of $\Delta
\alpha/\alpha$ is shown roughly in \cite{Dent}. In  following, We
try to further test the prediction of (\ref{alpha3}) in terms of
these $z$-dependent data of $\Delta \alpha/\alpha$. In order to
transfer the $t$-dependence of $\Delta \alpha/\alpha$ in
(\ref{alpha3}) to a $z$-dependence prediction,  a relation of $t-z$
is needed. For this aim, an appropriate cosmological model is
necessary sine the description of cosmical evolution is over the
$\mathcal{SR}_{cR}$ framework. The model considerations are follows:
1) The distance between the earth and QSO in the Beltrami reference
frame with origin of the earth (which is an inertial reference frame
in $\mathcal{SR}_{cR}$) $Q^1$ is caused by  comoving motion due to
the expansion of the Universe. And $Q^0/c=t$ is the comoving time;
2) The comoving time $t$ is determined by $\Lambda$CDM model
\cite{Lambda,Lambda1}. In this model, we have $t-z$ relation as
follows
\begin{equation}\label{La1}
t=\int_0^z{dz' \over H(z')(1+z')},
\end{equation}
where
\begin{eqnarray*}
\label{La2}H(z')&=&H_0\sqrt{\Omega_{m0}(1+z')^3+1-\Omega_{m0}},\\
\label{La3}H_0&=&100\;h\simeq 100\times0.705 km\cdot s^{-1}/Mpc,\\
\label{La4}\Omega_{m0}&\simeq &0.274.
\end{eqnarray*}
The $t-z$ relation is shown in Fig.(\ref{fig2}). Substituting this
relation into (\ref{alpha3}), we obtain desirous $z$-dependence
prediction of ${\Delta  \alpha \over \alpha} (z)$, where $R$ is free
parameter. By using observation data ${\Delta  \alpha \over \alpha}
(z=1.47)=-0.58\times 10^{-5}$, we get $R\simeq 2.73\times 10^{12}
ly$ (which is consistent with the estimation in \cite{Yan2}). Then
the theory predictions are ${\Delta  \alpha \over \alpha}
(z=0.65)=-0.24\times 10^{-5}$ and ${\Delta  \alpha \over \alpha}
(z=2.84)=-0.87\times 10^{-5}$, which are in agreement with the
corresponding data in \cite{Murphy03b} and \cite{Dent}. The results
are listed in table II, and the curve of ${\Delta  \alpha \over
\alpha} (z)$ is shown in Fig.(\ref{fig3}).

\begin{table}
\caption{ Time variations of $\Delta \alpha/\alpha$: The first two
columns are quoted from \cite{Dent}. Eq. (\ref{alpha3}) with
 $R\simeq 2.73\times 10^{12} ly$, and the $\Lambda$CDM model's $t-z$ relation (\ref{La1}) are used. }
\tabcolsep 0.1in
\begin{tabular}{c c c c} \hline \hline
average of redshift $\langle z\rangle$ & $(\Delta
\alpha/\alpha)_{expt}$ & epoch $t$ & theory prediction of
(\ref{alpha3}) \\ \hline 0.65 & $(-0.29\pm0.31)\times 10^{-5}$ &
$6.04$Gyr &$-0.24\times 10^{-5}$
\\
1.47 & $(-0.58\pm0.13)\times 10^{-5}$ & $9.29$Gyr &$-0.58\times
10^{-5}$
\\
2.84 & $(-0.87\pm0.37)\times 10^{-5}$ & $11.39$Gyr &$-0.87\times
10^{-5}$
\\\hline \hline
\end{tabular}
\end{table}

The comparison concludes that the theory predictions of
(\ref{alpha3}) agree with the observation data.

The  observation data of time variation of $m_p/m_e$ have also be
reported \cite{m-m}. However, there are no yet data of time
variation of $m_e$ up to now. Further experimental check to these
predictions is expected.

\section{Summery and discussions}

\noindent In this paper, we have solved the de Sitter special
relativistic ($\mathcal{SR}_{cR}$-) Dirac equation of hydrogen in
the earth-QSO framework reference by means of the adiabatic
approach. The  effects of de Sitter space-time geometry described by
Beltrami metric are  taken into account. The
$\mathcal{SR}_{cR}$-Dirac equation of hydrogen turns out to be
 a time dependent quantum Hamiltonian system. We have provided an
explicit calculation to examine whether the adiabatic approach to
deal with this time-dependent system is eligible. Since the radius
of de Sitter sphere $R$ is cosmologically large, it  makes the
time-evolution of the system is so slow that the adiabatic
approximation legitimately works with high accuracy. Finally, we
revealed that all those facts yield important conclusions that the
electromagnetic fine-structure constant, the mass of electron and
the Planck constant are of variation with time. Eqs. (\ref{alpha3})
(\ref{alpha2}) and (\ref{hme1}) describe the variations.

As is well known that the solutions of quantum mechanics equations
for atom of hydrogen played important roles for promoting the
development of quantum physics in the past century and achieved
several very great successes, such as to reveal dynamic bases
fundamentally for the Bohr level of hydrogen, the periodic law of
elements, the fine structure of atomic spectra and so on. The
studies on the fine structure of hydrogen specifically reveal the
effects due to combination of the quantum mechanics and the special
relativity $\mathcal{SR}_c$. The predictions were verified by
experiments. What were further promoted in this paper is that the
cosmology effects are involved via the solution of the
$\mathcal{SR}_{cR}$-quantum mechanics equation for hydrogen. The
time-variations of the fine structure constant and the mass of
electron are of cosmologic effects in atomic spectra.  Obviously,
the studies presented in this paper is different from other
theoretic considerations for this matter  from other insights, e.g.,
Kaluza-Klein theories \cite{Wu1}\cite{Wu2}, superstring
\cite{Macda}, accelerating Universe and dark energy \cite{Fujii},
etc.

There are several methods to study the time-variation of fine
structure constant experimentally. Among them, the observations of
absorption spectra of distant interstellar clouds
\cite{webb06,Murphy03b,Murphy03a,murphy01b,murphy01a,webb01,webb99}
are the most direct verification to the predictions of this paper.
Our result of the time variation of fine structure constant  are
consistent with the observations. This fact indicates that the
effects of de Sitter special relativity become visible at the cosmic
space-time scale (i.e., the distance $\geq 10^9 ly$). At that scale
de Sitter special relativity is more reliable than Einsteinian
special relativity, and the latter is the former's approximation for
the distance $<< R$. Finally we address that further experimental
tests are expected.

\section*{Acknowledgments}

I would like to thank Professors Zhe Chang, Chao-Guang Huang,
Han-Ying Guo, Jilang Jing, Qi-Keng Lu (K.H. Look), Yu Tian, Zhan Xu,
Huan-Xiong Yang, Yang Zhang for
 helpful discussions. Especially, I am grateful to Professor
Chao-Guang Huang for his critical comments and suggestion. I also
thank Wei Huang, Xiao-Dong Li for information discussions.
 The work is supported in part by National
Natural Science Foundation of China under Grant Numbers 10975128,
and and by  the Chinese Science Academy Foundation under Grant
Numbers KJCX-YW-N29.

\begin{appendix}
\section{Electric Coulomb Law in QSO-Light-Cone Space}

\noindent Let's derive (\ref{potential-B1}). We start with
(\ref{potential-B}), i.e.,
\begin{equation}\label{A-1}
-B^{ij}(Q)\pa_i\pa_j\phi_B(x)\hskip-0.07in=\hskip-0.06in\left(\nabla^2+{c^2t^2\over
R^2}{\pa^2\over \pa (x^1)^2}\right)\hskip-0.06in
\phi_B(x)\hskip-0.06in=\hskip-0.06in{-4\pi e\over
\sqrt{-det(B_{ij}(Q))}}\delta^{(3)}(\mathbf{x}),
\end{equation}
where $ -B^{ij}(Q)=\eta^{ij}-{c^2 t^2\over R^2}
\delta_{i1}\delta_{j1}+\mathcal{O}(R^{-4})$ has been used, and
$B_{ij}$ were given in (\ref{metric5}). Expanding (\ref{A-1}), we
have
$$
\left[{\pa^2\over \pa (x^1/[1+{c^2t^2\over 2R^2}])^2} +{\pa^2\over
\pa (x^2)^2} +{\pa^2\over \pa (x^3)^2}\right]\phi_B(x)=
-4\pi\left(1+{c^2t^2\over 2R^2}\right)e\delta(x^1)
\delta(x^2)\delta(x^3),$$ and further
\begin{eqnarray*}
&&\left[{\pa^2\over \pa (x^1/(1+{c^2t^2\over 2R^2}))^2} +{\pa^2\over
\pa (x^2)^2} +{\pa^2\over \pa (x^3)^2}\right]\phi_B(x)\\
&=&-4\pi e\delta({x^1\over 1+{c^2t^2\over 2R^2}})
\delta(x^2)\delta(x^3).
\end{eqnarray*}
Setting  $\td{x}^1\equiv x^1/(1+{c^2t^2\over 2R^2})$, the above
equation becomes
\begin{equation}\label{A-01}\left[{\pa^2\over \pa
(\td{x}^1)^2} +{\pa^2\over \pa (x^2)^2} +{\pa^2\over \pa
(x^3)^2}\right]\phi_B(x) =-4\pi e\delta(\td{x}^1)
\delta(x^2)\delta(x^3).
\end{equation}
 Then the solution is
$\phi_B(x)=e/r_B$ with
\begin{eqnarray}\label{rB}
\nonumber r_B&=&\sqrt{(\td{x}^1)^2+(x^2)^2+(x^3)^2}\\
\nonumber &=&\left((1-{c^2t^2\over 2R^2})^2
(x^1)^2+(x^2)^2+(x^3)^2\right)^{1/2}\\
&\simeq & r\left(1-{c^2t^2(x^1)^2\over 2R^2r^2} \right).
\end{eqnarray}
Therefore, we have
\begin{equation}\label{A-3}
\phi_B={e\over r_B}\simeq {e\over r}\left(1+{c^2t^2(x^1)^2\over
2R^2r^2} \right),
\end{equation}
which is eq.(\ref{potential-B1}) in the text.

\section{Adiabatic approximative wave functions in
$\mathcal{SR}_{cR}$-Dirac equation of hydrogen}

\noindent Now we derive the wave function of (\ref{wave1}) in the
text. We start with eq.(\ref{Dirac8}), i.e.
\begin{equation}\label{B1}
 i\hbar\pa_t\psi= H(t)\psi=[H_0(r_B, {e})+H'(t)]\psi,
\end{equation}
where
\begin{eqnarray}
\label{B021} H(t)&=&H_0(r_B,{e})+ H'(t),\\
 \label{B2} H_0(r,{e})&=& -i\hbar c
\vec{\alpha}\cdot\nabla_B + \mu c^2\beta
-{{e}^2\over r_B}\\
\label{B3} H'(t)&=&-\left({c^2t^2\over 2R^2}\right) H_0(r_B,
{\sqrt{2}}{e}).
\end{eqnarray}
Suppose the modification of $H(t)$ along with the time change is
sufficiently slow, the system could be quasi-stationary in any
instant $\theta$. Then, in the Shr\"{o}dinger picture, the
quasi-stationary equation of $H(\theta)$
\begin{equation}\label{B5}
H(\theta)U_n(\mathbf{x}, \theta)=E_n(\theta) U_n(\mathbf{x}, \theta)
\end{equation}
can be solved. By (\ref{B021}) (\ref{B2}) (\ref{B3}) and
$t\rightarrow \theta$, the solutions are as follows (similar to
eq.(\ref{solution1}) in text)
\begin{eqnarray}\label{B06}
E_n(\theta)\equiv E_{n_r,K}(\theta)&=&\mu_\theta
c^2\left[1+{\alpha_\theta^2 \over
(\sqrt{K^2-\alpha_\theta^2}+n_r)^2} \right]^{-1/2} \\ \nonumber
n_r&=&1,\;2,\;\cdots,~~~ |K|=(j+1/2)=1,\; 2,\; 3,\; \cdots,
\end{eqnarray}
where
\begin{eqnarray}
\label{B07} \mu_\theta &=&\left(1-{c^2\theta^2 \over 2R^2}\right)\mu
\\
 \label{B08} \alpha_\theta &=& \left(1-{c^2\theta^2 \over
3R^2}\right) {{e}^2\over \hbar c}\\
\label{B071} (\rm{and}\;\; \hbar_\theta &=&\left(1-{c^2\theta^2
\over 2R^2}\right)\hbar\; )
\end{eqnarray}
 The complete set of
commutative observable is $\{H,\; K,\;\mathbf{j}^2,\;j_z\}$, so that
we have
\begin{equation}\label{B09}
U_n(\mathbf{x},\theta)=\psi_{n_r, K, j,
j_z}(\mathbf{r}_B,\hbar_\theta, \mu_\theta, \alpha_\theta),
\end{equation}
where $\mathbf{j}=\mathbf{l}+{\hbar\over 2} \mathbf{\Sigma},\;\hbar
K=\beta(\mathbf{\Sigma}\cdot \mathbf{l}+\hbar)$. $[U_n(\mathbf{x},
\theta)]$ is complete set and satisfies
\begin{equation}\label{B10}
\int d^3x U_n(\mathbf{x}, \theta)U_m^*(\mathbf{x},
\theta)=\delta_{mn},~~~~n=\{n_r, K, j, j_z\}.
\end{equation}
 Thus, the solution of time-dependent
Shr\"{o}dinger equation (or Dirac equation) (\ref{B1}) can expanded
as follows
\begin{equation}\label{B6}
\psi(\mathbf{x}, t)=\sum_n C_n(t) U_n(\mathbf{x}, t)
\exp\left[-i\int_0^t\omega_n(\theta)d\theta\right],~~~
\omega_n(\theta)={E_n(\theta)\over \hbar}.
\end{equation}
Substituting (\ref{B6}) into (\ref{B1}), we have
\begin{equation}\label{B7}
i\hbar\sum_n(\dot{C}_nU_n+C_n\dot{U}_n)\exp\left[-i\int_0^t\omega_n(\theta)d\theta
\right]=0.
\end{equation}
By multiplying
$U^*_m\exp\left[i\int_0^t\omega_m(\theta)d\theta\right]$ to both
sides of eq.(\ref{B7}), and doing integral to $\bf{x}$ by using
(\ref{B10}), we have
\begin{eqnarray}\label{B13}
\dot{C}_m+C_m\int d^3x U_m^*\dot{U}_m
&+&\sum_n\hskip0.01in'\;C_n\int d^3xU_m^*\dot{U}_n\exp
\left[-i\int_0^t
(\omega_n-\omega_m)d\theta\right]=0, \\
\nonumber &m&=1,2,3,\cdots
\end{eqnarray}
where $\sum'_n$ means that $n\neq m$ in the summation over $n$.
Noting (\ref{B10}), we have
\begin{equation}\label{B14}
\int \dot{U}^*_m U_m d^3x+\int U_m^*\dot{U}_md^3x=0,
\end{equation}
and hence
\begin{equation}\label{B15}
\int U_m^*\dot{U}_md^3x=i\beta
\end{equation}
is purely imaginary number. Denoting
\begin{equation}\label{B16}
\alpha_{mn}=\int U_m^*\dot{U}_n d^3x,~~~~{\rm
and}~~~~\omega_{nm}=\omega_n-\omega_m,
\end{equation}
then eq.(\ref{B13}) becomes
\begin{eqnarray}\label{B17}
\dot{C}_m+i\beta C_m +\sum_n\hskip0.01in'\;C_n\alpha_{mn}\exp
\left[-i\int_0^t \omega_{nm}d\theta\right]=0.~~~~ &m&=1,2,3,\cdots
\end{eqnarray}
To further simplify it, we set
\begin{equation}\label{B18}
V_n(\mathbf{x},t)=U_n(\mathbf{x},t)\exp\left[-i\int_0^t\beta_n(\theta)d\theta\right],
\end{equation}
then
\begin{equation}\label{B19}
\psi(\mathbf{x}, t)=\sum_n C_n^{\;'}(t) V_n(\mathbf{x}, t)
\exp\left[-i\int_0^t\omega_n(\theta)d\theta\right],
\end{equation}
where
$C_n^{\;'}(t)=C_n(t)\exp\left[i\int_0^t\beta_n(\theta)d\theta\right]$,
and
\begin{equation}\label{B20}
\dot{C}_m^{\;'}(t)=[\dot{C}_m+i\beta_mC_m(t)]\exp\left(i\int_0^t\beta_n(\theta)d\theta\right)
\end{equation}
Substituting (\ref{B20}) into (\ref{B17}), we finally get
\begin{eqnarray}\label{B21}
\dot{C}_m^{\;'}+\sum_n\hskip0.01in'\;C_n^{\;'}\alpha_{mn}\exp
\left[-i\int_0^t \omega'_{nm}d\theta\right]=0.~~~~ &m&=1,2,3,\cdots
\end{eqnarray}
where
\begin{equation}\label{B22}
\omega'_{mn}=\omega'_n-\omega'_m,~~~~\omega'_n={1\over
\hbar}E_n+\beta_n.
\end{equation}
Now let's solve (\ref{B21}). Firstly, we derive $\alpha_{mn}$. By
(\ref{B5}), we have
\begin{equation}\label{B23}
{\pa H\over \pa t}U_n +H\dot{U}_n=\dot{E}_nU_n+E_n\dot{U}_n.
\end{equation}
By multiplying $U^*_m $ and doing integral over $\bf{x}$, we have
\begin{eqnarray}\nonumber
\int U^*_m\dot{H}U_nd^3x&+&\int U^*_mH\dot{U}_nd^3x=E_n\int
U^*_m\dot{U}_nd^3x \\
\label{24}{\rm
i.e.,}~~~~~\dot{H}_{mn}+E_m\alpha_{mn}&=&E_n\alpha_{mn},
\end{eqnarray}
so that
\begin{equation}\label{B25}
\alpha_{mn}=\int U^*_m\dot{U}_nd^3x={1\over
E_n-E_m}\dot{H}_{mn},~~~~ m\neq n.
\end{equation}
Therefore eq.(\ref{B21}) becomes
\begin{eqnarray}\label{B26}
\dot{C}_m^{\;'}+\sum_n\hskip0.01in'\;C_n^{\;'}{\dot{H}_{mn}\exp
\left(-i\int_0^t \omega'_{nm}d\theta\right)\over \hbar
\omega_{nm}}=0.~~~~ m=1,2,3,\cdots
\end{eqnarray}
Suppose in the initial time the system is in $s$-state, i.e.,
$C_n(0)=C_n^{\;'}(0)=\delta_{ns}$. To adiabatic process,
$\dot{H}(t)\rightarrow 0$, then the 0-order approximative solution
of eq.(\ref{B26}) is
\begin{equation}\label{B27}
[C_m^{\;'}(t)]_0=\delta_{ms}.
\end{equation}
Substituting (\ref{B27}) into (\ref{B26}), we get the first order
correction to the approximation
\begin{eqnarray}\label{B28}
[\dot{C}_m^{\;'}]_1={-\dot{H}_{ms} \over \hbar \omega_{ms}}
\exp\left(-i\int_0^t \omega'_{ms}d\theta\right)=0,~~~~ m\neq s.
\end{eqnarray}
Since the dependent on time $t$ of $U_n(t)$ is weak for adiabatic
process, eq.(\ref{B15}) indicates $\beta_n$ is small, and by
(\ref{B22}), we have $\omega'_{ms}\approx \omega_{ms}$. Then, from
(\ref{B28}), the first order correction to the solution is
\begin{eqnarray}\label{B29}
[C_m^{\;'}]_1={\dot{H}_{ms} \over i\hbar \omega_{ms}}
\left(e^{i\omega_{ms}t}-1\right),~~~~ m\neq s.
\end{eqnarray}
Substituting (\ref{B28}) (\ref{B29}) into (\ref{B19}) and neglecting
$\beta_n$, we get the wave function as follows
\begin{equation}\label{B30}
\psi(\mathbf{x}, t)\simeq  U_s(\mathbf{x}, t)e^{-i{E_s t\over
\hbar}}+\sum_{m\neq s}{\dot{H}_{ms} \over i\hbar \omega_{ms}}
\left(e^{i\omega_{ms}t}-1\right)U_m(\mathbf{x},
t)e^{\left(-i\int_0^t{E_m(\theta)\over \hbar}d\theta\right)}.
\end{equation}
By using eqs.(\ref{B09}), (\ref{B08}), (\ref{B07}), (\ref{B071}), we
finally obtain the desired results
\begin{equation}\label{wave11}
\psi(t)\simeq\psi_s(\mathbf{r}_B,\hbar_t,
\mu_t,\alpha_t)e^{-i{E_s\over \hbar}t}+\sum_{m\neq
s}{\dot{H}'(t)_{ms} \over i\hbar
\omega_{ms}^2}\left(e^{i\omega_{ms}t}-1\right)\psi_m(\mathbf{r}_B,\hbar_t,
\mu_t,\alpha_t)e^{\left(-i\int_0^t{E_m(\theta)\over
\hbar}d\theta\right)},
\end{equation}
where
\begin{eqnarray}
\label{alpha1} \alpha_t &=& \left(1-{c^2t^2\over
2R^2}\right){\alpha},~~{\rm with}~~{\alpha}={{e}^2\over \hbar c},\\
\label{mu1} \mu_t&=& \left(1-{c^2t^2\over 2R^2}\right)\mu,\\
\label{mu1} \hbar_t&=& \left(1-{c^2t^2\over 2R^2}\right)\hbar.
\end{eqnarray}
They are just the equations (\ref{wave1}), (\ref{alpha}),
(\ref{hme2}) and (\ref{hme1}) in the text.

\end{appendix}

%\end{CJK*}

\end{document}